\documentclass[a4paper,twoside]{article}
\newcommand{\affil}[1]{$^{\rm #1}$}
\usepackage[authoryear]{natbib}
\bibpunct{(}{)}{;}{a}{}{,}
\usepackage{graphicx}
\date{} 
\title{\large\bf\flushleft The SkyMapper Telescope and The Southern Sky Survey}
\author{\parbox{\textwidth}{\flushleft
\vspace{-0.5cm}
{\it S.\ C.\ Keller\affil{A}\affil{B}, B.\ P.\ Schmidt, M.\ S.\ Bessell, P.\
  G.\ Conroy, P.\ Francis, A.\ Granlund, E.\ Kowald, A.\ P.\ Oates, T.\
  Martin-Jones, T.\ Preston, P.\ Tisserand, A.\ Vaccarella, M.\ F.\
  Waterson}\\
\vspace{0.4cm}
{\small \affil{A}\,Research School of Astronomy and Astrophysics, Australian
  National University, Cotter Rd, Weston, ACT 2611, Australia}\\
{\small \affil{B}\,Email: stefan@mso.anu.edu.au}}}
\begin{document}
\twocolumn[
\begin{changemargin}{.8cm}{.5cm}
\begin{minipage}{.9\textwidth}
\vspace{-1cm}
\maketitle
%
%
\small{\bf Abstract:}
This paper presents the design and science goals for the SkyMapper
telescope. SkyMapper is a 1.3m telescope featuring a 5.7 square degree field\--of\--view Cassegrain imager commissioned for the Australian National
University's Research School of Astronomy and Astrophysics.  It is located at
Siding Spring Observatory, Coonabarabran, NSW, Australia and will see first light in late 2007.

The imager possesses 16k$\times$16k 0.5 arcsec pixels. The primary scientific
goal of the facility is to perform the Southern Sky Survey, a six colour and
multi-epoch (4 hour, 1 day, 1 week, 1 month, 1 year sampling) photometric
survey of the southerly 2$\pi$ steradians to $g\sim$23 mag. The survey will
provide photometry to better than 3\% global accuracy and astrometry to better
than 50 mas. Data will be supplied to the community as part of the Virtual
Observatory effort. The survey will take five years to complete.


\medskip
\medskip
\end{minipage}
\end{changemargin}
]
\small

\section{Project Overview}
SkyMapper is amongst the first of a new generation of dedicated, wide-field
survey telescopes to start operation in the coming five years. It is now
feasible to tessellate the field of view of a low f-number telescope with CCDs
and achieve areal coverage comparable to that obtained with photographic
Schmidt cameras. The all-sky photographic surveys of the Palomar and UK
Schmidt telescopes \citep{rei91,hol74} are currently unsurpassed in the
optical. Such photographic surveys have several shortcomings however, they do
not provide photometry to better than 0.2 mag.\ or astrometry to better than
0.5 arcsecs.

Several groups are now actively pursuing digital multi-colour surveys of the sky, notably Pan-Starrs \citep[operational 2007;][]{kai02} and the Large Synoptic Survey Telescope\citep[first light in 2012;][]{cla04}. The only survey already in progress is the Sloan Digital Sky Survey \citep[SDSS;][]{yor00}. SDSS have mapped close to $\pi$ steradians of the northern sky in five colours. The gamut of scientific applications for SDSS data spans properties of the asteroid population \citep{ive01} to the discovery of the most distant quasar \citep{fan01}. In the infrared, VISTA \citep{vistaref} plans to survey the southern sky starting in early 2007. Other large aperture survey instruments are planned such as ESO's VST \citep[first light 2007;][]{cap04} and the Lowell Observatory's Discovery Channel Telescope \citep[first light 2009][]{seb04} with narrower fields of view designed for targeted deeper surveys.

SkyMapper will perform the multi-colour, multi-epoch Southern Sky Survey
(S3). Reflective of our science goals we have devised a filter set for the S3
that is optimised for stellar astrophysics. We have sought a filter set that
best discriminates the important stellar parameters of effective temperature,
surface gravity and metallicity. This is not to the detriment of non-stellar
science as these areas are well served by a series of broadband filters.

The SkyMapper focal plane will feature a mosaic of 32 2k$\times$4k CCDs coupled to high speed device controllers for rapid, low noise readout. The facility will operate in an automated manner and require minimal operational support. On site there is a scheduling and data quality assurance system. Data is transferred via a high speed link to the Australian National University's Supercomputing Facility where the data reduction pipeline resides. SkyMapper will see first light in 2007.

\section{The SkyMapper Telescope}

\subsection{Site}
\begin{figure*}
\begin{center}
\includegraphics[scale=0.5, angle=0]{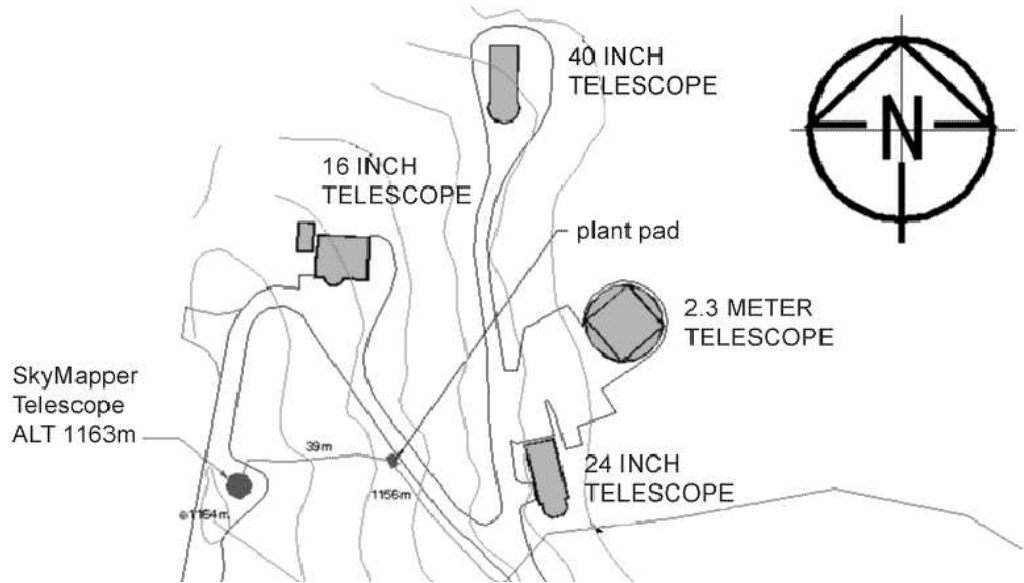}
\caption{Plan of the SkyMapper site on Siding Spring Mountain.}\label{figsite}
\end{center}
\end{figure*}

The SkyMapper telescope is located 20m north-east of the summit of Siding Spring Mountain, near Coonabarabran, NSW (altitude 1169m). Figure \ref{figsite} shows a plan of the observatory site. To minimise the facility's impact on the seeing obtained with the telescope, large heat sources such as power supplies, computers and instrumentation within the enclosure are cooled by chilled water supplied from an equipment pad 40m downhill from the telescope.

\subsection{Enclosure}
\begin{figure}
\begin{center}
\includegraphics[scale=0.4, angle=0]{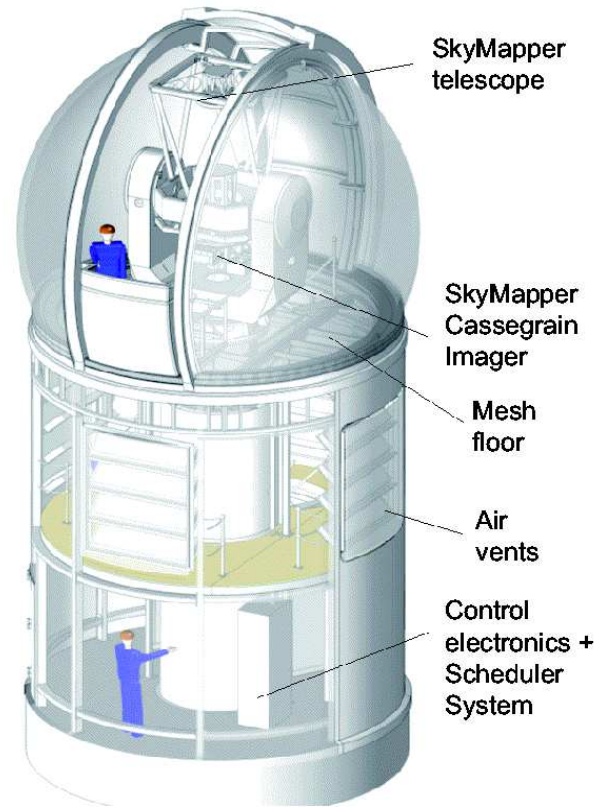}
\caption{The SkyMapper enclosure.}\label{figenclosure}
\end{center}
\end{figure}

The SkyMapper enclosure is 6.5m in diameter and 11m high with three internal levels as shown in Figure \ref{figenclosure}. The first level houses the electronics and control computers and is thermally isolated from the rest of the building. Four sets of vent shutters are set into the walls of the second level. When observing these shutters enable passive ventilation of the observing space above via the mesh floor of the observing space. The vent shutters will have controlled opening angles to flush the observing space while not introducing excessive wind buffeting. 

During the day the volume of levels 2 and 3 will be sealed and actively cooled to the expected ambient temperature at the start of observing for the night ahead. The interior of the dome will be finished matt black to minimise scattered light reaching the detector.

The SkyMapper enclosure features internal and external arrays of weather sensors. These will prevent operation in inclement weather. Should mains power fail, there will be sufficient power in the enclosure's UPS to safely shutdown the facility.

Lightning is a significant risk for the facility as Siding Spring Mountain is exposed to severe summer thunderstorms (the vicinity receives approximately 3 strikes per km$^{2}/$yr). The mountain top is largely outcropping bedrock of high resistivity. Our lightning mitigation consists of a series of six air terminals that connect to a radial ground plane of buried copper braid that radiates 20m from the enclosure. In addition, the telescope control computer will receive information from a commercial lightning warning service and will close the enclosure in the case of an impending thunderstorm.

\subsection{Site Conditions}
\label{sectseeing}

The prevailing astronomical conditions at the site are an important input to the design of the project.  We have examined the weather logs of the Anglo-Australian Telescope (AAT) from 2000-2005\footnote{C. McCowage private communication}. Siding Spring Observatory's (SSO) weather patterns show little seasonal variation, with an average of 2110 hours (64\%) of the time useable throughout the year, of which 1135 hours (35\%) is photometric.

\begin{figure}
\begin{center}
\includegraphics[scale=0.4, angle=0]{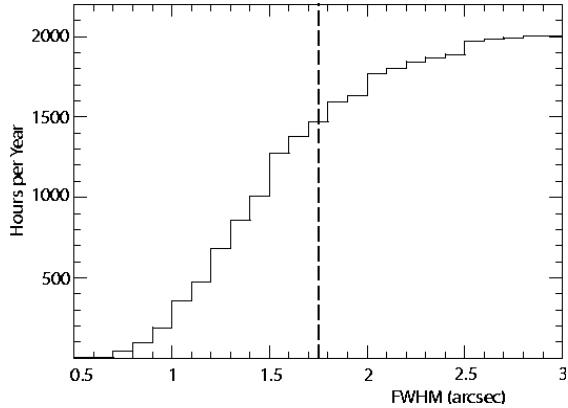}
\caption{Seeing at Siding Spring derived from logs of the Anglo-Australian Telescope.}\label{figseeing}
\end{center}
\end{figure}

Using DIMM measurements \citep{woo95} the median free air seeing at SSO is
approximately 1.1 arcsecs. This assumes the turbulence is not near the ground
and therefore the actual seeing may be somewhat worse. This is borne out by the logs of the AAT which are summarised in Figure \ref{figseeing}. This shows 68\% of the time the seeing is less than 1.75 arcsecs.

\begin{figure}
\begin{center}
\includegraphics[scale=0.3, angle=0]{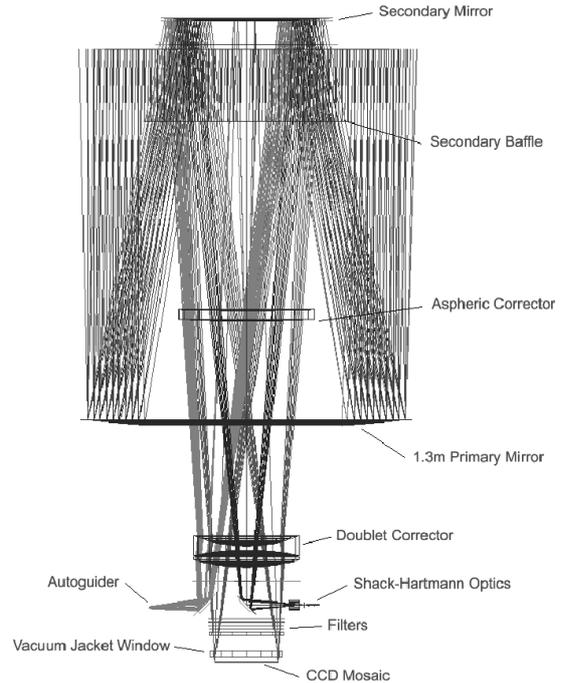}
\caption{SkyMapper optics.}\label{figoptics}
\end{center}
\end{figure}

\subsection{Optical design}

The SkyMapper telescope optical system was designed by Electro Optics Systems,
Australia. It features a 1.33m primary mirror and a relatively large 0.69m
secondary. This results in a collecting area equivalent to an unobstructed aperture of 1.13m. The telescope is a modified Cassegrain design, optimised for wide-field operation between 340nm and 1000nm. The design in shown in Figure \ref{figoptics}.

The primary mirror is of Astrosital glass ceramic fabricated by LZOS, Russia. The secondary mirror (sourced from SAGEM, France) is carried on an actuated hexapod system that allows for tip-tilt, x\&y and telescope focus movement. The telescope control system uses the hexapod to automatically adjust for gravity induced flexure and focus change due to temperature.

The primary and secondary mirrors possess protected aluminium coatings. This increases the longevity of the reflective surface, particularly in the UV. The three transmissive corrector elements are single-layer anti-reflection coated and composed of fused silica to maximise the UV throughput of the instrument.

The EOS SkyMapper telescope is a compact, stiff and rugged structure with high mechanical natural frequencies. We expect the telescope to point to $<\pm$3 arcsec and track to 0.5 arcsec RMS for 5 minutes in winds of $<$5m/s. Both of the telescope principal axes are driven by direct on-axis DC ring motors without intervening gearboxes and direct on-axis incremental encoders. The system is therefore free of backlash.

\begin{table}
\begin{center}
\caption{SkyMapper Imager Parameters}\label{tabimager}
\begin{tabular}{ l  r }
\small Telescope working f/ratio & \small f/4.7878\\
 \small Telescope Focal Plane Scale & \small 0.0302mm/arcsec\\
 \small Detector Pixel Projection & \small 0.5 arcsec/pixel\\
 \small Number of Pixels in Mosaic & \small268 435 456\\
 \small Mosaic Dimensions & \small 256.34$\times$258.75mm\\
 \small Mosaic Field of View & \small 2.373$^{o}$$\times$2.395$^{o}$\\
 \small Mosaic Fill Factor & \small 91.05\%\\
 \small Sky Coverage w. Fill Factor & \small 5.68 square degrees\\
\end{tabular}
\end{center}
\end{table}

\section{The SkyMapper Imager}

The SkyMapper Imager has been designed in-house at RSAA (see \citet{gra06} for more details). Due to the large focal plane, the instrument is comparatively large and heavy for the size of telescope. To achieve an optimal outcome, the designs of the telescope and imager have proceeded in parallel, with close cooperation between the RSAA and Electro-Optic Systems design teams. Important imager parameters are summarised in Table \ref{tabimager}.

\begin{figure}
\begin{center}
\includegraphics[scale=0.4, angle=0]{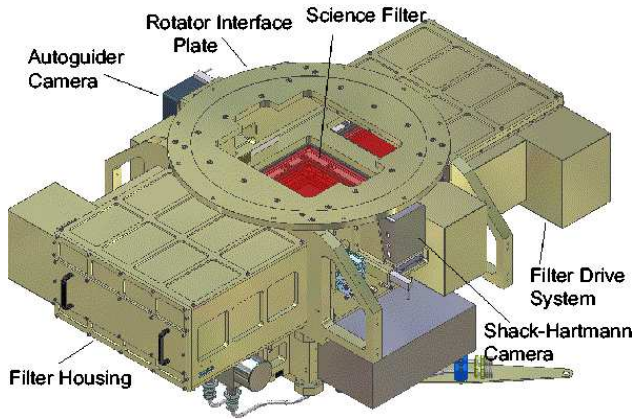}
\caption{The SkyMapper Cassegrain Imager as seen from above.}\label{figimager}
\end{center}
\end{figure}

\begin{figure}
\begin{center}
\includegraphics[scale=0.4, angle=0]{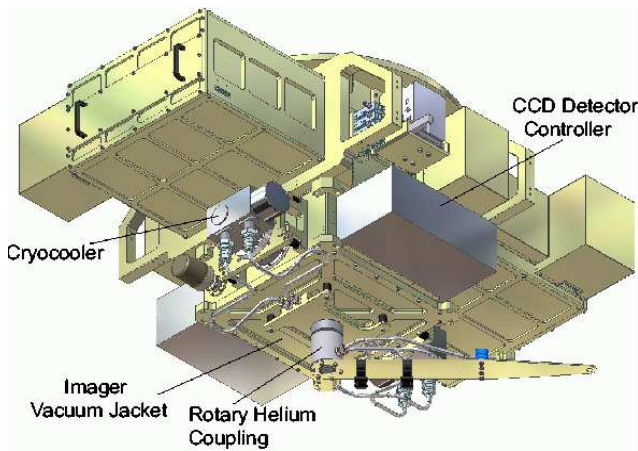}
\caption{The SkyMapper Cassegrain Imager as seen from below.}\label{figimager2}
\end{center}
\end{figure}

Figure \ref{figimager} shows the imager instrument from above looking down on the rotator interface plate. The autoguider and science filters can be seen through the beam aperture. Also seen is the Shack-Hartmann system that is used for automatic collimation and focusing. Figure \ref{figimager2} shows the underside of the imager. The Imager vacuum jacket is at the centre, flanked by closed-cycle helium cryocoolers and twin detector controllers. The instrument weighs $\sim$300 kg.

A large mechanical shutter has been fabricated by the University of Bonn, Germany. The shutter is located below the filter stack and above the vacuum jacket window. The shutter is composed of two blades that form a moving slot of variable width to provide uniform exposure of the focal plane. The shortest exposure is about 1ms and exposure homogeneity is 0.3\% in a 100ms exposure \citep{rei04}.

SkyMapper has six interchangeable filters, each 309\-mm square and up to 15mm
thick. Table 4 lists the specifications of the filter set for the S3. Our
motivation for the choice of filters is detailed in Section
\ref{secfilters}. The filter set is constructed of coloured glass where
possible. Coloured glass filters show better throughput homogeneity across the
focal plane than can be currently achieved with multi-layer interference
filters. The filter glass was sourced from Macro Optica, Russia and Schott,
Germany. Only the $r$ and $i$ band filters have additional short wavepass
coatings to define the bandpass. An optimised anti-reflection coating will be
applied to each filter by Optical Surface Technologies, USA.

The Imager vacuum jacket carries the CCD mosaic as shown in Figure \ref{figvacuumjacket}. The vacuum jacket window is 25mm thick fused silica with a broadband anti-reflection coating. A steady flow of dry air is passed over this window to prevent condensation.

\subsection{The SkyMapper Imager CCD Mosaic}

\begin{figure}[t]
\begin{center}
\includegraphics[scale=0.4, angle=0]{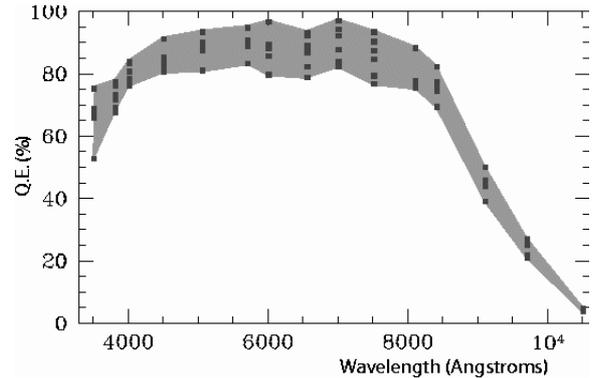}
\caption{The spectral response of SkyMapper science CCDs. The shaded area
  encloses the minimum and maximum measured response for a set of six devices.}\label{figqes}
\end{center}
\end{figure}

The SkyMapper Imager CCD mosaic is a 4 x 8 array of E2V CCD44-82 detectors. Each
CCD detector has 2048 x 4096, 15 micron square pixels. The deep depletion devices are back-illuminated and 3-side buttable. They possess excellent quantum efficiency from 350nm-950nm (see Figure \ref{figqes}), near perfect cosmetics, and low-read noise.

All 32 CCDs are carried on a 10mm thick Invar carrier plate. The surface of the carrier plate has been machined to 10$\mu$m flatness and on to this surface the precision pads of the E2V devices are mounted. Our aim for focal plane flatness is to match the single pixel geometric depth of field in the f/4.78 beam of $\pm32\mu$m.

\begin{figure}
\begin{center}
\includegraphics[scale=0.4, angle=0]{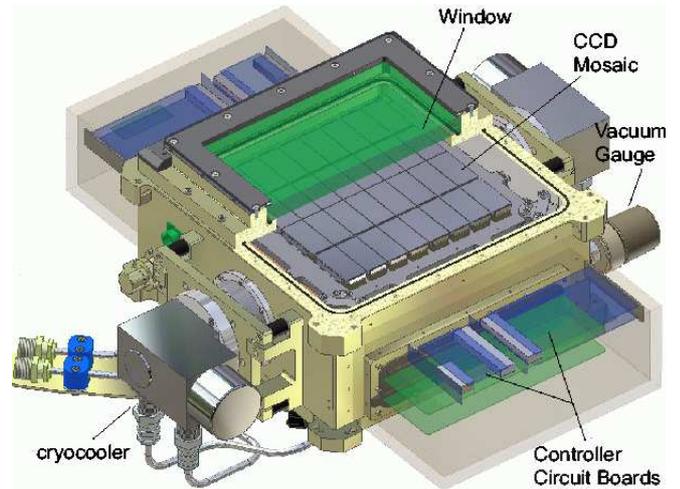}
\caption{The SkyMapper Cassegrain Imager vacuum jacket showing the CCD mosaic.}\label{figvacuumjacket}
\end{center}
\end{figure}

A portion of the top of the CCD detector mosaic assembly can be seen in Figure \ref{figvacuumjacket}. The inner two rows of CCDs are butted together with a 1.5mm gap between the rows and a 0.5mm gap between columns. A larger gap exists between the centre and outer rows of about 4mm. This assembly gives a filling factor of 91\%.

\subsection{CCD Controllers}
The SkyMapper Cassegrain Imager will utilise two customised versions of the newly developed 16-channel STARGRASP controllers\footnote{see http://www.stargrasp.org/} developed for Pan-STARRS by Onaka and Tonry et al. from the University of Hawaii.  A hybrid 300MHz PowerPC CPU FPGA is used in the controller for digital signal processing. Each controller also possesses 256MB of embedded memory for buffering the high data flows. The controllers will have integrated Gigabit Ethernet optical-fibre ports for fast readout, which will enable the SkyMapper Imager to easily meet its requirement of reading out the entire focal plane in $<$20s (450kpix/s). It is our goal to read out the focal plane in ~10s, i.e. ~1Mpix/sec. This new detector controller technology will also reduce the size and weight of the controller hardware substantially and increase reliability over legacy systems. The controllers are to be attached to either side of the Detector Vacuum Enclosure (see Figure \ref{figvacuumjacket}).

Both detector controllers will be connected to a single pixel server computer
that transfers the data to local storage. The pixel server is commanded by the
Computerised Instrument Control And Data Acquisition software system
\citep[CICADA:][]{you99} developed by RSAA for all observatory
instruments. The RSAA's Telescope Automation and Remote Observing System
\citep[TAROS:][]{wil05} in turn controls CICADA to configure the instrument and
acquire the required exposure.

\section{Science Goals}
\label{sectscigoals}
The science that can be done with a comprehensive survey such as the Southern
Sky Survey cannot be fully described or predicted. Data from previous all-sky
surveys has been used in thousands of papers and important discoveries
continue to be based upon them decades after their completion. In the
following section we outline a series of scientific questions that we have
identified SkyMapper will have high impact in addressing. These science goals define the requirements for SkyMapper's Southern Sky Survey.

\subsection{What is the distribution of solar system objects beyond Neptune?}
The existence of quiescent comet-like objects orbiting the Sun in a region
beyond Neptune was first proposed by \citet{leo30} following the discovery of
Pluto. The objects in this region remained undiscovered until 1992 when the
first Trans-Neptunian Object (TNO) was discovered \citep{jew93}. Subsequent
searches have, to date, uncovered more than 1000 TNOs. TNOs are believed to be
debris left over from the formation of the solar system. The distributions of
their orbits and masses provide important clues to the process of outer
planet formation and to the origin and nature of comets. An all-sky survey
will overcome the observational biases that exist in current work based around
the ecliptic. Little is known about any larger TNOs, especially those
on scattered orbits that take them out of the ecliptic. Models of
\citet{ken99} predict many large TNOs, some with masses larger than Pluto. The
recent discovery of 2003UB313, a scattered object with a diameter slightly in
excess of Pluto \citep{bro06}, demonstrates the possibilities for future
discovery.

The requirements for this science program are as follows. By going to
$r\sim$21.5 over the entire sky we stand to discover $\sim$2000 TNOs
\citep{tru01}. The motions are typically a few arcsec per hour at
opposition. To distinguish nearby objects from distant ones, at least 3
observations are required. We envisage two epochs separated by $\sim$4hrs on
the first night with a third epoch 1-3 days later. A fourth epoch after one
month will be sufficient for extrapolation of the orbit to the next
year. Astrometric accuracy is not particularly stringent: $<\pm$0.2 arcsec
will suffice.

\subsection{What is the history of the young\-est stars in the solar neigh\-bour\-hood?}
In the last two decades it has been possible to definitively discern coeval,
comoving associations of young stars within the solar neighbourhood. By
looking at the space motions and spatial distribution of young objects
\cite{kas97} revealed the TW Hya association. With the advent of all-sky X-ray
and deep proper motion catalogues further close young associations were found
($\eta$ Cha; \cite{mam99}, Tucana-Horologium; \cite{tor00} and
\cite{zuc00}). These associations of young stars have been recently ejected
from the molecular clouds of their birth, most likely within the Sco-Cen
association, a massive region of star formation in the southern Milky Way
\citep{zuc04}.

Because these young stars are unobscured and close to the Sun, they are ideal
targets for high resolution imaging of protoplanetary disks with Spitzer,
SOFIA and IR adaptive optics imagers. These objects offer us a view of how
stars and planetary systems form, one not obscured by the material from which
they formed. While several comoving associations have been found from the
Hipparcos and Tycho catalogues, these surveys are limited to the brightest
members of the cluster luminosity function.

A key requirement for this science program is to have good proper motions for objects across the sky as well as accurate colour selection down to $r$=18. This will enable us to select nearby candidates for planetary disk searches. Relative astrometry to 50mas over a five year baseline will provide proper motions to 4masyr$^{-1}$ accuracy. This is sufficient to find relevant objects to 120pc.

\subsection{What is the shape and extent of the dark matter halo of our galaxy?}
Dark matter is the dominant gravitational component of our Universe. We have
few clues to its nature. One clue is its distribution. It is now clear that
dark matter in the cores of galaxies does not produce central cusps of
material as predicted by theory \citep{nav00}. Even less is
known about how dark matter is distributed in the outer regions of
galaxies. For the Milky Way we would like to know the extent and shape of the
dark matter: is it spherical, flattened or triaxial? The answers to these
questions place constraints on the nature of dark matter and the formation of
our galaxy.

Visible dynamical tracers must be used to probe the distribution of dark
matter in the galactic halo. Perhaps the best such tracers are RR Lyrae stars
but such stars are rare: approximately one per 10 square degrees beyond 50kpc
and one per 100 square degrees beyond 100kpc \citep{ive01}. Blue horizontal
branch stars are 8-10 times more common, but require a good surface gravity
discriminant to separate from the bulk of higher surface gravity main sequence
stars and blue stragglers in the field. Intrinsically luminous F and G giants
can be seen to greater distances but are as rare as RR Lyraes and similarly
need a good surface gravity indicator.

Requirements for this program are a series of at least two exposures over
short ($\sim1$ week) time frame to discriminate RR Lyraes. 90\% of RR Lyrae
stars can be recovered with a $g$$-$$r$ colour to 0.1mag precision and 6
epochs in $g$ with a precision of 0.1mag. Blue horizontal stars can be
discriminated on the basis of colour information with a gravity indicator to
distinguish contaminating blue stragglers and main-sequence A-type stars (see
Section \ref{secfilters}). Selection of blue horizontal branch stars and
giants requires photometry accurate to $\sim$0.03mag.

\subsection{Extremely metal-poor stars: how did our Galaxy evolve?}
What were the characteristics of the first stars, when did they light up the
Universe, and how did they form into galaxies like the Milky Way? The
formation of the first stars is a key moment in the history of the Universe
yet it lies approximately 13 billion years in the past \citep{nao06}, beyond
the reach of modern-day telescopes. However, our galaxy holds a fossil record
of the first generations of stars. Since the Big Bang gave rise to a universe
of hydrogen, helium and an admixture of light elements, successive generations
of stars have synthesized and returned to the interstellar medium the
$\sim$4\% by mass of heavier elements that comprise, for instance, the Sun. By
using metallicity as a proxy for age, the most metal-poor stars are candidates
for the first generation of stars. These candidates will then require
follow-up spectroscopy to determine such quantities as radial velocities and
fine abundance analysis.

Requisite for this program is accurate ($\sigma<0.03$ mag) multi-colour
photometry from the near-UV to the red for stars between
12$-$18$^{th}$magnitude. Large sky coverage is essential as such stars are
extremely rare. The choice of filters should provide optimal determination of
effective temperature, surface gravity and metallicity.

\subsection{High redshift QSOs: when did the first stars in the Universe form?}
We know that within 1.5 billion years after the Big Bang, some galaxies had
formed; this is evidenced by galaxies observed to z$\sim$6. Even at this epoch
most of the intergalactic medium was ionized as evidenced from the lack of
continuum absorption redward of Lyman-$\alpha$ (the Gunn-Peterson
effect; \citet{gun65}). It must be concluded that some objects must have
existed earlier that produced sufficient UV flux to ionise nearly all the
baryonic matter in the Universe.

The SDSS located a series of z$>$5.8 quasars including the most distant object
at z=6.42 \citep{fan01}. The most distant QSO does seem to show continuous
Lyman-$\alpha$ absorption \citep{bec01}. This is our first glimpse of the
interface between the ionised intergalactic medium we see locally and the
neutral material before. The description of this surface tells us about the
physics of the era of reionisation \citep{wyi04}.

So far three high redshift objects from the SDSS show the Gunn-Peterson effect. By covering the entire southern sky we stand to take this to a dozen or more such objects. Requirements for this program are survey sensitivities matching those of SDSS. Multiple epochs are essential for good cosmic ray rejection.

The union of the above requirements provides the defining (most difficult to
achieve) constraints of the survey. They are summarised in Table
\ref{tabreqs}.

\begin{table*}[ht]
\begin{center}
\caption{Southern Sky Survey defining requirements}\label{tabreqs}
\begin{tabular}{|l|c|c|c|c|c|c|}
\hline & $u$ & $v$ & $g$ & $r$ & $i$ & $z$\\
\hline Sensistivity & $\sigma=0.03$ mag& $\sigma=0.03$ mag& $\sigma=0.1$
mag & $\sigma=0.1$ mag & $3\sigma$ & $7\sigma$\\
& $u<$20 & $v_{s}<20$ & $g=21.2$ & $r=21.0$ & $i=22.5$ & $z=20.5$\\
& & & & & detection & detection\\
\hline Cadence & & & \small 6 epochs & \small 6 epochs & \small 3 epochs for&
\small 3 epochs for\\
& & &\small hours to &\small hours to & \small cosmic-ray&\small cosmic-ray\\
& & &\small years &\small years & \small rejection &\small rejection \\
\hline Systematic&\multicolumn{6}{|c|}{0.03 mag} \\
\hline Astrometry&\multicolumn{6}{|c|}{50 mas over 3 years}\\
\hline
\end{tabular}
\end{center}
\end{table*}

\subsection{Non-Survey Science}
{\bf{Planetary transits}} - SkyMapper allows monitoring of a large field of
  view to high stability (relative flux to better than 0.01mag) making it a
  highly competitive facility for this purpose.

{\bf{Supernovae}} - We are proposing to use time that does not meet the
  survey's seeing and sky background constriants to undertake a supernova
  survey. Such a survey will provide continuous coverage of 1250 square
  degrees of sky, and discover approximately 100 SN Ia to z$<$0.085 per
  year. In addition to allowing a systematic exploration of the statistical
  properties of Type Ia supernovae, it will also serve to populate the SN Ia
  Hubble Diagram in this low redshift regime. Similarly, for core collapse
  supernovae we will increase statistics with which to explore the energetics
  and nucleosynthetic production of these explosions. The supernova survey
  will lead to the detection of many hypernovae with which to explore the
  relationship between these objects and Gamma-Ray Bursts.

\section{Southern Sky Survey Design}
The requirements of Table \ref{tabreqs} are met by the design of the
Southern Sky Survey that we now detail. Our science goals
emphasize the need for wide sky coverage. We have defined 4069 fields
covering the 2$\pi$ steradians of the southern sky. 

\subsection{Survey Coverage and Cadence}

To meet our science goals we must capture variability on short and long time
scales. Since most variable astronomical objects have time scales for
variability ranging from hours (RR Lyraes, asteroids), months (supernovae,
long period variables) and years (QSOs, stellar parallax and proper motions)
our survey design will ensure that each field is observed in each filter on
six epochs with approximately the following cadence, t = 0, $+$4 hours, $+$1 day, $+$1 week, $+$1 month, $+$1 year.

The six epochs will be spatially offset slightly from each other according to
a dither pattern. This will ensure that chip gaps and defects do not occult
the same area of sky. At completion, 90\% of the sky will be
covered at least five times and 100\% imaged three or more times.

The survey is to be completed in five years to realize its full
potential. Together with the weather constraints detailed in Section
\ref{sectseeing}, we have defined our exposures to be 110 seconds
(plus 20 second overhead between expsures) in each filter. Two data releases
are planned: a first data release when 3 epochs in each filter have been
obtained and calibrated and a second when the complete set of 6 epochs in each
filter have been obtained. The expected survey depths in 1.5 arcsec seeing for
a 5$\sigma$ detection in AB magnitudes are given in Table \ref{tabdepth}.

\begin{table}[ht]
\begin{center}
\caption{SkyMapper Main Survey depth in AB magnitudes (signal-to-noise of 5 - 110 second exposures).}\label{tabdepth}
\begin{tabular}{|l|c|c|c|c|c|c|}
\hline & $u$ & $v$ & $g$ & $r$ & $i$ & $z$\\
\hline \small 1 epoch & \small 21.5 & \small 21.3& \small 21.9 & \small 21.6 & \small 21.0 & \small 20.6\\
\small 6 epochs & \small 22.9 & \small 22.7& \small 22.9 & \small 22.6 & \small 22.0 & \small 21.5\\
\hline
\end{tabular}
\end{center}
\end{table}

\subsection{Optimised for Stellar Astro\-physics - filter set design}
\label{secfilters}
\begin{figure}[ht]
\begin{center}
\includegraphics[scale=0.4, angle=0]{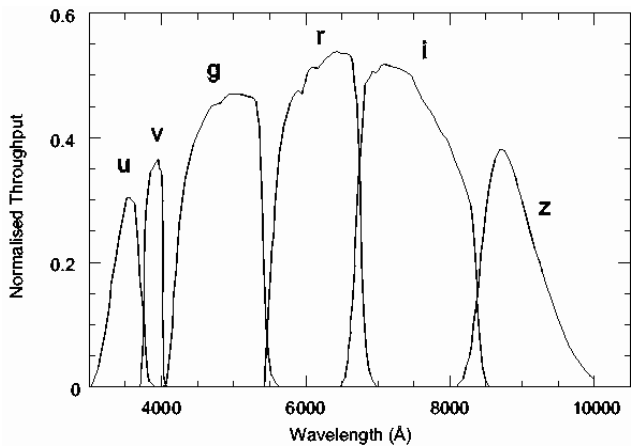}
\caption{The expected throughput ex-atmosphere of the filter set for the
  Southern Sky Survey.}\label{figthroughput}
\end{center}
\end{figure}

The filter set we have designed for S3 provides both inter-operability with
existing systems such as SDSS and key information for the science goals
discussed in Section \ref{sectscigoals}. The specification of the filter set
is given in Table \ref{tabfilters}. In Figure \ref{figthroughput} we show the
expected throughput of the SkyMapper Telescope and Cassegrain Imager. We have
taken as our basis the Gunn \& Thuan $griz$ filter set \citep{thu76} utilized
by SDSS\footnote{http://www.astro.princeton.edu/PBOOK/camera/
camera.htm}. When compared to the SDSS system the SkyMapper system features
twice the throughput in $u$ and three times the throughput in $z$, albeit with
our smaller aperture. The cleanly separated $i$ and $z$ filters are optimal
for discovery of high redshift QSOs (Lyman alpha in $z$ equates to a redshift
of greater than 5.8). Photometric redshifts for z$<$0.5 galaxies are
facilitated by broadband photometry over the optical spectrum.

A large part of the science potential of SkyMapper is bound to the study of
stellar populations. It was important, therefore, for us to consider how we can
best characterize stellar populations using broad to intermediate band
photometry. To do this we computed the synthetic colours of stars spanning
metallicity, surface gravity, effective temperature and extinction from
\citet{castelli06} model atmospheres. For any given point in the colour-colour space, the number of synthetic models within an error ellipsoid provides the measure of uncertainty in the derived quantities. These simulations were performed as a function of photometric uncertainty and filter choice.

For G stars and hotter the main discriminating power is derived from the near-UV. We incorporate a Str\"omgren $u$ filter and an intermediate band filter similar to DDO38, we term $v$, that covers the spectrum from 3670-3980\AA. These two filters are able to break the degeneracy between surface gravity and metallicity as illustrated by the following important applications.

\begin{table}[ht]
\begin{center}
\caption{SkyMapper filter pass bands}\label{tabfilters}
\begin{tabular}{|c|c|c|c|}
\hline Filter & 50\% Cut-on & 50\% Cut-off & FWHM \\
& edge (\AA) &  edge (\AA) & (\AA) \\
\hline $u$    & 3250 & 3680 & 430\\
 $v$ & 3670 & 3980 & 310\\
 $g$ & 4170 & 5630 & 1460\\
 $r$ & 5550 & 7030 & 1480\\
 $i$ & 7030 & 8430 & 1400\\
 $z$ & 8520 & 9690$^a$ & 1170$^a$\\
\hline
\end{tabular}
\medskip\\
$^a$ $z$ filter is unblocked at red end, the red edge is limited by the CCD
sensitivity cut off in the near-IR.
\end{center}
\end{table}

\subsubsection{Application 1: Blue Horizontal Branch Stars}
\begin{figure}[ht]
\begin{center}
\includegraphics[scale=0.45, angle=0]{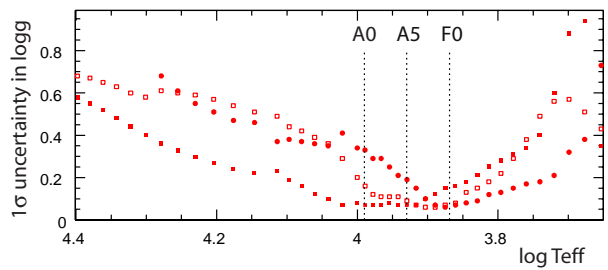}
\caption{The one sigma uncertainty in surface gravity derived under
  photometric uncertainties of 0.03 mag from the SkyMapper filter
  set. Solid squares are log g=4.5, open squares log g=3.5 and circles
  log g=2.0. Dashed vertical lines show the temperatures corresponding to
  spectral types on the main sequence.}\label{figdellg}
\end{center}
\end{figure}

In Figure \ref{figdellg} we consider the uncertainty in the derived stellar
surface gravity as a function of temperature for stars of log g = 2.5 and
4.5 with photometric uncertainties of 0.03 magnitudes per filter. For hot O and B stars the intensity of the Balmer jump feature is not significantly affected by changes in surface gravity. Consequently we have limited ability to discern surface gravity for such stars. Similarly, for stars cooler that 5000K (G6-7V) the Balmer jump disappears as a spectral feature. However, for A type stars we expect to determine the surface gravity to $\sim10$\%. This range in temperature is inhabited by blue horizontal branch stars (BHBs) - standard candles for the Halo.

\begin{figure}[ht]
\begin{center}
\includegraphics[scale=0.45, angle=0]{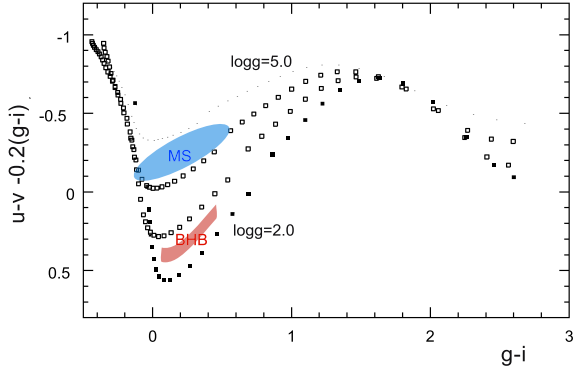}
\caption{$u$$-$$v$ vs. $g$$-$$i$ for stars of solar metallicity and a
  range of surface gravity. Blue horizontal branch stars are well separated
  from main-sequence and blue straggler stars.}\label{figumvs}
\end{center}
\end{figure}

A sightline through the halo inevitably contains local disk main-sequence
stars and blue stragglers in the temperature range of interest. However, these
contaminants are of significantly higher surface gravities than those of
BHBs. The $u$ and $v$ filters measure the Balmer Jump and the
effect of H$-$ opacity which increases with surface gravity. As can been seen
from Figure \ref{figumvs}, the $u$$-$$v$ colour index discriminates between
BHBs and their contaminants. The separation is 0.5mag in colour. On the basis
of our photometric selection we will be able to discern a sample of BHBs to a
distance of 130kpc with less than 5\% contamination.

\subsubsection{Application 2: Extremely metal-poor Stars}

\begin{figure}[ht]
\begin{center}
\includegraphics[scale=0.45, angle=0]{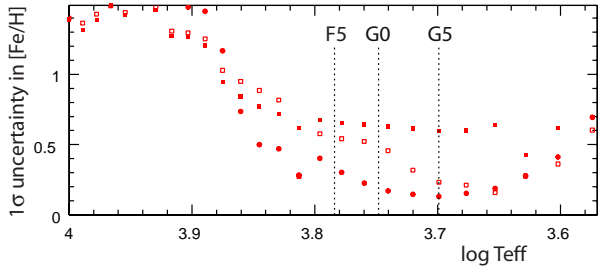}
\caption{The one sigma uncertainty in metallicity derived under photometric
  uncertainties of 0.03 mag from the SkyMapper
  filter set. Solid squares represent [Fe/H]=-4 models, open squares
  [Fe/H]=-2 and circles [Fe/H]=0.}\label{figdelfeh}
\end{center}
\end{figure}

Amongst cooler stars, the $u$ and $v$ filters capture the magnitude of
metal line blanketing, a general suppression of the continuum blueward of
$\sim$4000\AA. In Figure \ref{figdelfeh} we show the uncertainty in
derived metallicity as a function of temperature for stars of metallicity
[Fe/H] = -4 from the S3 filter set. 

We are able to determine [Fe/H] best at high metallicity and the uncertainty
rises to $\pm$0.7dex at [Fe/H]=-4. At extremely low metallicities the density
of stars drops by a factor of ten for every factor of ten drop in
metallicity. Consequently, the challenge in finding extremely metal-poor stars
(EMPs) is to keep the very small subset of interesting objects clean from
photometric outliers. In this regard, the S3 filter set provides improved
separation over existing filter sets between the bulk of halo stars of
[Fe/H]$>-2$ and those of [Fe/H]$<-4$. This provides us a distinct advantage in
isolating extremely metal-poor stars from photometric colours.

\begin{figure}[ht]
\begin{center}
\includegraphics[scale=0.45, angle=0]{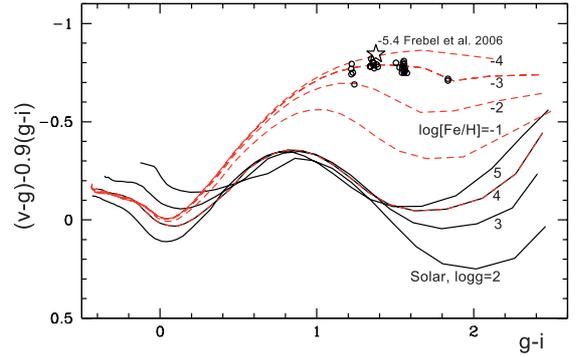}
\caption{$v$$-$$g$ vs. $g$$-$$i$ for stars of solar metallicity and a
  range of surface gravity (solid lines) and for log g=4 and metallicities
  to [Fe/H]=-4. Overlaid are the computed colours of HE1327-2326 \citep{fre05}
  and the sample of extremely metal-poor stars from \cite{cay04}.}
\label{figvsmg}
\end{center}
\end{figure}

Our ability to discern metallicity is not clearly expressed in a colour-colour
plot as the vector of grad metallicity is not closely aligned with any
colour. The best correspondence between grad metallicity and our colours is
found with the $v$$-$$g$ colour shown in Figure \ref{figvsmg}. We see that
$v$$-$$g$ has little dependence on surface gravity from early F to mid G;
however this breaks down as we enter the cooler K-type range. The evolved extremely metal-poor stars do not possess temperatures as cool as K0 and those cooler than K0 and still on the main-sequence are sufficiently lacking in luminosity to make their numbers small over the survey volume. 

We have calculated the number of halo stars in the survey
brighter than $g$=18.1 (where we reach a signal-to-noise ratio of 33 in the
first data release) by integrating the Bahcall and Soniera galactic model
\citep{bah80} over the survey area with galactic latitude greater than 15
degrees. We then scale by the results of the Hamburg ESO Survey \citep{chr03}
to find the number of stars with [Fe/H]$<$-4. There are on order of a thousand
such stars, including a hundred of metallicities less than -5. Follow-up will
be tractable; to recover 90\% of the stars of [Fe/H]$> -4$ will require
intermediate resolution spectroscopy of $\sim$2600 stars. High signal-to-noise
8m spectra will be required on the distilled sample of extremely metal-poor
stars for detailed elemental abundance studies.

\section{Photometric Calibration}
\label{sectphotom}

To calibrate the S3 to the required global accuracy of 0.03 magnitudes we will perform a shallow survey of the entire southern sky under photometric conditions (the Five-Second Survey). Exposure times of approximately five seconds in each filter will enable us to obtain photometry for stars from 8.5 to 15.5 magnitudes. The Five-Second Survey will consist of a set of at least three images of each field in all filters. The network of standards will provide the photometric and astrometric catalogue to which the deeper Main Survey images will be anchored. With all-sky coverage, the Five-Second Survey will enable the Main Survey images (and any other images taken with the telescope) obtained in non-photometric conditions to be calibrated.

The major impediment to the derivation of accurate photometry from wide field of view instruments is the influence of scattered light. \cite{mag04} studied the systematic errors in the flat-field structure of the CFHT12K imager. By dithering a standard star sequence over the CCDs of the mosaic they were able to form a map of the photometric residuals (observed - expected) over the focal plane (a photometric superflat). In the case of the CFHT12K imager systematic errors in the flatfield amount to 5\% peak-to-peak: around half due to pixel scale change across the camera, another half due to the effect of scattered light. The scattered light component arises from ambient light scattered off surfaces within the enclosure and telescope reaching the detector. We have sought to minimise the scattered light by finishing the interior surfaces matt black, however some scattered light will inevitably remain. 

Forming a photometric superflat is a direct way to remove the signature of
scattered light. During commissioning, we will establish a series of six
calibration fields at declination $\sim$-25$\deg$ (to avoid the zenith keyhole
of the telescope) around the sky. By observing the calibration fields with a
series of dithers and rotator angles we will be able to characterise and
correct for departures from photometric flatness as a function of position on
the detector array. These fields will also serve as our secondary standard
star fields. During Five-Second Survey operations we will acquire standard
star fields every 90 minutes. At such time the two highest fields will be
observed, thus providing a range of airmasses from $\sim1$ to 1.8.

With full coverage of right ascension a photometric ``ring closure solution'' can be formed (see for example \citet{pel06}). The ring closure solution ensures that the differential flux ratios between calibration fields in all bandpasses are very accurately determined. This, finally, leaves the zeropoints of the flux in each filter to be determined. We will define the absolute zero point of the SkyMapper system from the Walraven photometric system (Pel \& Lub ibid) and the associated stars in the Next Generation Spectral Library\footnote{Gregg et al. http://lifshitz.ucdavis.edu/\-~mgregg/\-gregg/\-ngsl/\-ngsl.html}. To do this we will first use the Walraven photometry to accurately zeropoint the spectrophotometric standards. We will then use the spectrophotometric standards and with our photometric bandpasses, to derive absolute zeropoints. Initially our photometric bandpasses will be those measured by a laboratory monochromometer convolved with the CCD spectral response and adjusted to match the observations of standard stars. Ultimately, we plan to use a monochromatic flat + NIS standard \citep{stu06} to characterise the response of our system.

\section{Astrometric Calibration}
\label{sectastro}

Our astrometric calibration uses as its basis the UCAC\-2 astrometric catalogue \citep{zac04}. The calibration has two stages. Firstly, UCAC2 stars in common to Five-Second images are used to define an image world coordinate system (WCS). The UCAC2 provides between 40 and several hundred stars per SkyMapper CCD. Our derived positional uncertainties are better than 50 milliarcsecs for bright stars. The Main Survey images are then astrometrically calibrated against the stars of the Five-Second Survey. 

We implement the ZPN\footnote{Zenithal/azimuthal polynomial projection see \cite{cal04}} coordinate representation. This representation utilizes a radial term to characterise distortions across the image and a simple linear transformation to remove the effects of rotation, scale, and any shear caused by refraction. Since most distortion is radial from the optical axis, the ZPN system is able to linearise a distorted optical plane with a single term.  Linearising the telescope system in this way provides robust coordinate transformations across the images with no difficulties in CCD corners or in images with few astrometric standards.

Any systematic astrometric offsets will be applied via a lookup table after processing a large portion of the survey. These are not expected, but would be difficult to accurately quantify if they depend on the rotator angle of the telescope. Differential atmospheric refraction will be corrected for once the colours of each object are determined. Over five years and 36 epochs, it will be possible to derive proper motions to $\pm$4mas/yr.

\begin{figure*}
\begin{center}
\includegraphics[scale=0.55, angle=270]{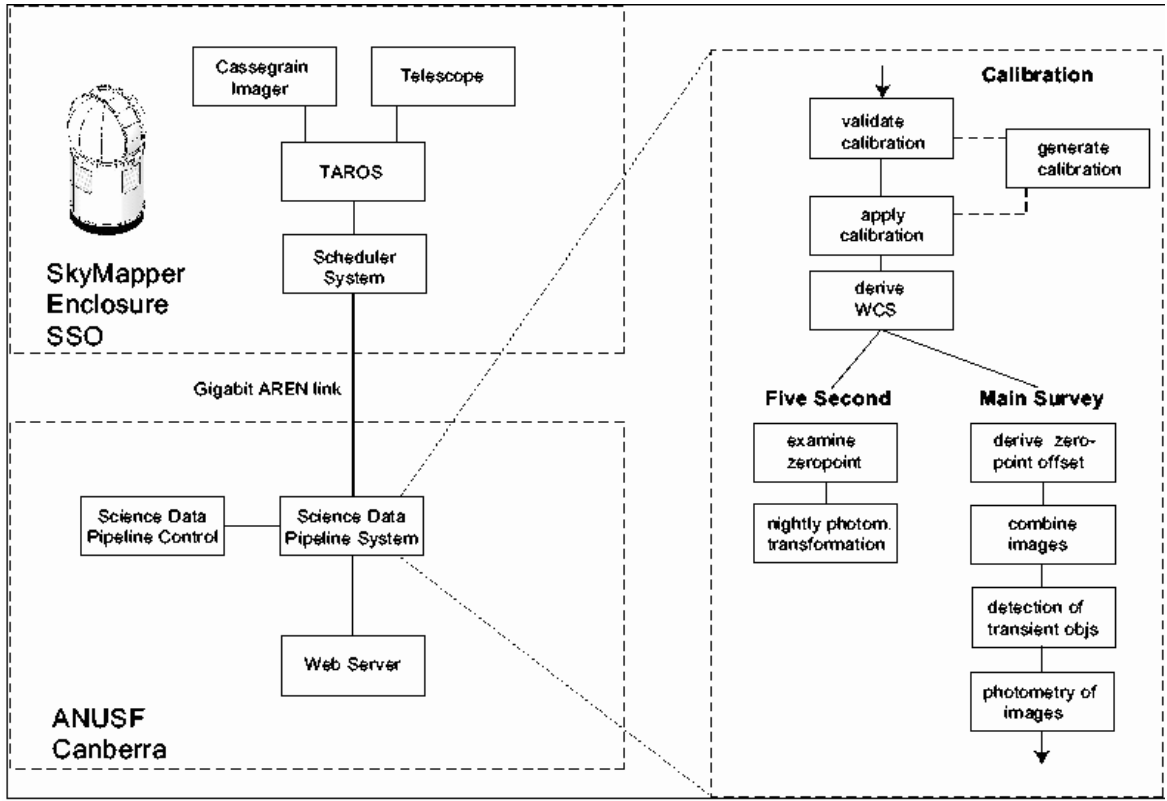}
\caption{A system level view of the SkyMapper Scheduler and Science Data Pipeline.}\label{figsystem}
\end{center}
\end{figure*}

\section{SkyMapper Data Acquisition \& Reduction}
The acquisition and reduction of SkyMapper data is managed by the Scheduler and Science Data Pipeline Systems respectively. A schematic system level diagram is given in Figure \ref{figsystem}. We now discuss the two systems in detail.

\subsection{The Scheduler System}
The on-site Scheduler System is responsible for scheduling the required calibration and science data. In the course of a typical night, first bias frames will be acquired then sky flatfields for as many filters as time permits. As each exposure is completed, the data is checked for quality and system changes. In the case of the bias frames we check that the bias levels are nominal. This is a simple check of the health of the detector system. Before astronomical twilight a focusing sequence will be performed. Alternatively, when available, the Shack-Hartmann system will be used to determine the telescope focus.

At the end of evening astronomical twilight, Sky\-Mapper will begin the task of acquiring S3 images. The required instrument configuration is supplied by the Scheduler System to the instrumentation via the TA\-R\-OS interface. At detector readout the next instrument configuration is issued by the Scheduler System to the telescope in order to minimise latency. The CCD devices are read out to a 5TB RAID array and the local disk. The copy of the image on local disk is used for data quality checks so as not to impede the writing of data to the RAID array.

The Five-Second Survey holds top priority in sched\-uling as this data must be obtained under photometric conditions. A manually set flag tells the Scheduler
System if the night is likely to be photometric based on predicted weather conditions from the Australian Bureau of Meteorology MLAPS model. If this flag is set, the scheduler will commence taking any required (and possible) Five-Second Survey data. After writing the acquired data to disk the Scheduler
System performs a check of the photometric zeropoint and the surface brightness in one CCD image. Rapidly fluctuating zeropoints or zeropoints below a nominal value indicate the presence of cloud.

Should the conditions become non-photometric during the night the Scheduler System will switch to the Main Survey. The Main Survey can continue under uniform diminished transparency such as might occur with light cirrus. If the transparency drops below a threshold then the data is flagged of poor quality and the field is made available for rescheduling. Due to time constraints for quality checks we do not attempt to monitor the uniformity of transparency across all 32 CCDs of a field within the Scheduler System. This check is performed in the Science Data Pipeline described below.

Another critical aspect of image quality, the seeing, is also monitored by the Scheduler System. The Main Survey will have a seeing limit of $\sim$1.75 arcsecs (the 68th percentile of seeing). The remainder of the time will be given to a poor seeing program. In the case of the Five-Second Survey, an upper limit of
2.5 arcsecs will suffice before the poor seeing program is pursued.

Data is trickled throughout the night and proceeding morning to the ANU
Supercomputing Facility (ANU\-SF) via a gigabit link. A winter night of
Five-Second Survey data (a maximal constraint) can be transferred over several hours. At completion of observing the dome will close and the telescope will park. The enclosure air conditioning will aim for the next night's expected temperature as predicted by the MLAPS model.

\subsection{The Science Data Pipeline System} 
Data reduction is managed by the Science Data Pipeline System (SDPS) running at the ANUSF, Canberra. The SDPS has been designed specifically for S3 data. Figure \ref{figsystem} outlines the stages of the SDPS for data from the Five-Second and Main Surveys. The system consists of a series of modules that perform the required calibration, photometric standardisation, catalogue generation, photometry, and data delivery via the Internet.

Pipeline modules are implemented in C/C++ for numerically intensive processes, with Perl scripting to facilitate process flow. Pipeline progress is tracked, and results recorded, in a PostgreSQL database. Individual modules are queued for execution on the Australian Partnership for Advanced Computing SGI Altix
3700Bx2 cluster housed at ANUSF.

The SDPS stages are as follows. Firstly, for each completed night we check the nightly median bias again\-st the median bias generated from the bias frames obtained on many (up to 30) nights preceding the night in question. If the nightly median bias is outside nominal bounds the system attempts to generate a new median bias to apply. A similar sanity check is applied to the flatfields obtained. If a comparison of the nightly flatfields with the current authorized flatfield is outside bounds, such as might occur due to a change in the optical system of the telescope caused by, for example, the deposition of dust, the system will attempt to generate a new median flatfield to apply to this and subsequent nights' data. When flatfield change occurs, there may not be sufficient numbers of flatfields to be combined to form an authorized flatfield and several nights' worth of data may not be able to be processed until such time as sufficient flatfields are obtained and an authorized flatfield produced.

In addition to the bias and flatfield calibration files, fringe frames are generated for $i$ and $z$. Skyflats are generated from the median of science images in each band. Skyflats are produced to remove most of the effects of scattered light in the system. We then derive a photometric superflat for each filter to characterize the residual effects of scattered light. The photometric superflat is formed by imaging of our calibration fields (Section \ref{sectphotom}). We form a median image from the residuals between our reference photometry and that obtained from images after bias, flat, fringe and skyflat application. It is necessary to regenerate the fringe, skyflat and superflat whenever the flatfield changes as all are intimately related to the optical path. For quality control, the processes that generate calibration data require the astronomer to authorize the new calibration files before they are applied to any data.

A world coordinate system is defined for each CCD image as described in
Section \ref{sectastro}. Once the Five-Second Survey data are in place, we test uniformity of transparency across each Main Survey image by deriving the zeropoints of all CCDs and examining their deviations from nominal. Significant spatial variation of zeropoints indicates the presence of non-uniform cloud attenuation. Such an image is rejected from further processing.

At this stage the image is ready for extraction of photometry and
astrometry. The SDPS has two streams for the two data types, the Five-Second
and Main Surveys.

For each night of five-second data the zeropoints of all images are
considered. Excessive variation indicates the night was not photometric and
the data is discarded. Aperture photometry is performed on the images. We then
derive the nightly transformation equations from the photometry derived from
the calibration fields that are interleaved with our five-second
observations. We then apply the transformation equations to the data residing
in the five-second database.

Main Survey images are reduced when three (first data release) or six (second
data release) images of a field have been obtained in all six filters and
corresponding five-second data exists. Firstly, the zeropoint offset between
each image and the corresponding Five-Second survey photometry is found on a
chip by chip basis. The photometry for each image is then corrected for this
offset to draw the potentially non-photometric Main Survey data onto the
standard instrumental magnitude system.

We combine the set of images in each filter. In doing so, we mask saturated
stars and grow their associated bleeding, remove satellite trails and remove
cosmic ray tracks. Objects within the combined image are detected to 7
sigma. Transient objects will not be retained in the median frame. In order
to capture transients we mask the objects found from the median image in each
component image and then look for detections greater than 10 sigma. A final catalog is constructed as the union of all detections in the g,r,i,z bands, as well as any transient objects.

Photometry is performed upon the complete catalog of objects. For each object,
flux will be measured via PSF-fitting, aperture (over a range of apertures),
Petrosian, and Kron photometry\footnote{using modified Source Extractor code
\citet{ber96}}. In addition, photometry will also be forced to occur in fixed
position mode (useful at low SNR), and where the centroid of the object is
allowed to drift. Classification of object type will occur utilising Source
Extractor's neural network. Galaxy photometry is obtained in two modes:
firstly, using the galaxy shape as determined from the image with the most
signal and, secondly, using the shape parameters from each median filter
image. The resulting object positions, photometry and shape parameters are
stored in the S3 database.

The Five-Second Survey represents 75 terabytes worth of data, the Main Survey
another 150 terabytes. This data is retained, in raw and reduced forms, at the ANUSF's Mass Data Store, a robotic tape archive of PB size. It is estimated that the database of photometry for the Main Survey will amount to of order 5TB.

\section{Data Products}

The first SkyMapper data product will be the Five-Second Survey database. This will be released when a ring closure solution has been established for the calibration fields (see Section \ref{sectphotom}). These photometric data will be accessible from the survey website. The website will offer the ability to perform simple relational queries on the database of photometry.

We envisage two Main Survey data releases for each field; one when three
epochs in all filters have been obtained and a second when the final set of
six observations have been obtained. The data release will staged after
adequate quality control has been undertaken. In addition to database queries
of the Main Survey photometry, external users will have access to the reduced,
combined images.





\section*{Acknowledgments} 
SkyMapper is funded by the Australian National University and supported by
grants from the Australian Research Council. Weather information from the AAT
logs was kindly provided by Chris McCowage and Steven Lee of the AAO.


\end{document}